\documentclass[epsfig,graphics,twocolumn,floatfix,a4paper,showpacs]{revtex4}
\usepackage{amsmath,amsfonts,amssymb,graphicx,color,times,bbm,psfrag,dsfont}
\usepackage[english]{babel}
\usepackage[latin1]{inputenc}
\usepackage[T1]{fontenc}

\newcommand{\beq}{\begin{equation}}
\newcommand{\eeq}{\end{equation}}
\newcommand{\beqa}{\begin{eqnarray}}
\newcommand{\eeqa}{\end{eqnarray}}
\newcommand{\ket} [1] {\vert #1 \rangle}
\newcommand{\bra} [1] {\langle #1 \vert}
\newcommand{\braket}[2]{\langle #1 | #2 \rangle}

\DeclareMathOperator*{\argmax}{max}

\usepackage{graphicx}
\usepackage[mathscr]{eucal}
\usepackage{amsmath}
\usepackage{amssymb}
\usepackage{epstopdf}
\usepackage{epsfig}
\def\one{\ensuremath{\hbox{$\mathrm I$\kern-.6em$\mathrm 1$}}}

\begin{document} 

\title{Universal geometric entanglement close to quantum phase transitions}
\author{Rom\'an Or\'us}

\affiliation{School of Physical Sciences, The University of Queensland, 
QLD 4072, Australia}

\begin{abstract}

Under successive Renormalization Group transformations applied to a quantum state $\ket{\Psi}$ of finite correlation length $\xi$, there is typically a loss of entanglement after each iteration. How good it is then to replace $\ket{\Psi}$ by a product state at every step of the process? In this paper we give a quantitative answer to this question by providing first analytical and general proofs that, for translationally invariant quantum systems in one spatial dimension, the global geometric entanglement per region of size $L \gg \xi$ diverges with the correlation length as $(c/12) \log{(\xi/\epsilon)}$ close to a quantum critical point with central charge $c$, where $\epsilon$ is a cut-off at short distances. Moreover, the situation at criticality is also discussed and an upper bound on the critical global geometric entanglement is provided in terms of a logarithmic function of $L$.

\end{abstract}
\pacs{03.67.-a, 03.65.Ud, 03.67.Hk}
\maketitle

{\it Introduction and aims.-}  Quantum phase transitions at zero temperature play a key role in the occurrence of important collective phenomena in quantum many-body systems. In this respect, a considerable effort has been applied throughout the last few years towards a theory of entanglement in extended systems, such as quantum lattice systems and quantum field theories. In particular, there have been several attempts to generalize the $c$-theorem of Zamolodchikov, which implies that the entanglement properties of a one-dimensional quantum system are  somehow lost along Renormalization Group (RG) trajectories \cite{zam}. A deeper understanding of this theorem using tools from quantum information science has shown that, under consecutive RG transformations,  a translationally invariant quantum system in one spatial dimension may suffer from a monotonic loss of its amount of bipartite entanglement along the flow \cite{ent1, ent3}, which can be explained in terms of a set of majorization relations (the so-called \emph{ fine-grained entanglement loss along RG flows} \cite{fin1}). This has in part motivated the application of a number of renormalization group ideas to novel representations of quantum states \cite{TN}. 

However, and in spite of the above findings, for a generic quantum system it is not yet known {\it how close} can its quantum state be to a globally separable state after each RG transformation. 
In other words, the behavior along RG flows of the \emph{global} entanglement in extended quantum systems --- that is, the multipartite correlations that are shared by many different parties --- still remains unclear in many aspects. So far, studies of global entanglement in extended systems have only been done for very specific quantum lattice models (see e.g. \cite{global1, global2, global3}), and solid analytic results of wide generality are still missing.

This letter deals with the above situation by providing first explicit analytical derivations, in the case of quantum systems in one spatial dimension and invariant under translations, of the universal properties along RG flows of the \emph{global geometric entanglement per region of size $L$}, which we call $\mathcal{E}$ \cite{global,global3}. As we shall see, this measure of entanglement allows to quantify the fidelity between the quantum state and the closest separable state of contiguous blocks of size $L$ along the flow. More precisely, we establish that near criticality and for one-dimensional quantum systems of finite correlation length  $\xi$, the global geometric entanglement gets saturated when increasing the size $L$ according to
\beq
\mathcal{E} = \frac{c}{12} \log{\left(\frac{\xi}{\epsilon}\right)} ~~~~~~~~ L \gg \xi \gg \epsilon \ 
\label{equ2}
\eeq
close to a quantum critical point with central charge $c$ \cite{CFT}, where $\epsilon$ is a regularization parameter at short distances that coincides with the lattice spacing for lattice systems. The above relation implies a logarithmic divergence of the saturation value of $\mathcal{E}$ with the correlation length $\xi$ when criticality is approached, so that the quantum system experiences a loss of multipartite entanglement along RG flows that decrease $\xi$. Furthermore, the situation at criticality is also  discussed, for which we provide the upper bound
\beq
\mathcal{E} < \frac{c}{6} \log{\left(\frac{L}{\epsilon}\right)} ~~~~~~~~ L \rightarrow \infty \ 
\label{equ3}
\eeq
implying that the average geometric entanglement per block can not grow faster than a logarithmic function in the size $L$, in agreement with previous numerical estimations for bosonic and fermionic lattice models \cite{global3}.  

\begin{figure}[h]
\includegraphics[width=0.45\textwidth]{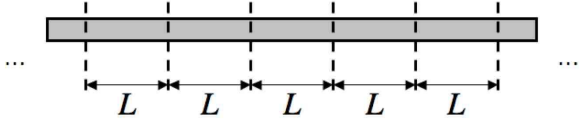}
\caption{An infinite one-dimensional quantum system is divided into different parties corresponding to contiguous blocks of size $L$.}
\label{fig1}
\end{figure}

{\it Global geometric entanglement.-} To introduce the measure of entanglement that we use throughout this paper we initially consider a pure quantum state of $N$ parties $\ket{\Psi} \in \mathcal{H} = \bigotimes_{i = 1}^N \mathcal{H}^{[i]}$, where $\mathcal{H}^{[i]}$ is the Hilbert space of party $i$. Our aim is to quantify the global multipartite entanglement of $\ket{\Psi}$. Following \cite{global}, this can be achieved by considering the maximum fidelity $|\Lambda_{{\rm max}}|$ between the quantum state $\ket{\Psi}$ and all the possible separable and normalized states $\ket{\Phi}$ of the $N$ parties, 
\beq
|\Lambda_{{\rm max}}| = \argmax_{\ket{\Phi}}|\braket{\Phi}{\Psi}| \ ,
\label{mma}
\eeq
which is related to the minimum distance between $\ket{\Psi}$ 
and the set of separable states. In order to have a measure of entanglement that is zero when $\ket{\Psi}$ is separable we take the natural logarithm, 
\beq
E(\Psi) = - \log{\left(|\Lambda_{{\rm max}}|^2\right)} \ .
\eeq
Here we will be interested in the above quantity per party, which has a well defined infinite-$N$ limit: 
\beq
\mathcal{E}_N = N^{-1} E(\Psi) ~ , ~~~~~~ \mathcal{E} \equiv \lim_{N \rightarrow \infty} \mathcal{E}_N \ . 
\eeq
In the case of a one-dimensional quantum system of infinite size we choose the different parties to be contiguous blocks of size $L$ as shown in Fig.(\ref{fig1}). In this setting, $\mathcal{E}$ corresponds to the global geometric entanglement per region of size $L$ \cite{global3}, and is the entanglement measure that consider in this work. 

{\it Loss of global entanglement along RG flows.-} Our aim now is to provide a derivation of Eq.(\ref{equ2}) in the introduction. 
The complete proof of that relation depends on a number of results that we have derived by using properties from linear algebra and conformal field theory. In the proof presented here, some strong technical details are avoided and referred correspondingly, always looking for the clarity of expression.

To start with, let us consider the quantum state $\ket{\Psi}$. For the moment, the total length of the system is assumed to be $NL$, and the limit $N \rightarrow \infty$ will be taken shortly. The first step in our derivation is to find a suitable decomposition of state $\ket{\Psi}$ as in \cite{guif1}: first, we consider the bipartition of the system $[1 : 2, \ldots, N]$ and compute the Schmidt decomposition of $\ket{\Psi}$,
\beq
\ket{\Psi} = \sum_\alpha \lambda^{[1]}_\alpha \ket{\tau^{[1]}_\alpha} \ket{\omega^{[2,\cdots,N]}_\alpha} \ .
\eeq
In the above equation, $\lambda^{[1]}_\alpha$ are the Schmidt coefficients, and $\ket{\tau^{[1]}_\alpha} , \ket{\omega^{[2,\cdots,N]}_\alpha}$ are the left and right Schmidt vectors respectively. Next, we find the Schmidt decomposition of quantum state $\ket{\omega^{[2,\cdots,N]}_\alpha}$ according to the bipartition $[2:3, \ldots, N]$, 
\beq
\ket{\omega^{[2,\cdots,N]}_\alpha} = \sum_\beta \lambda^{[2]}_\beta \ket{\tau^{[2]}_{\alpha \beta}} \ket{\omega^{[3,\cdots,N]}_\beta} \ ,
\eeq
so that state $\ket{\Psi}$ reads
\beq
\ket{\Psi}  = \sum_{\alpha, \beta}  \lambda^{[1]}_{\alpha} \lambda^{[2]}_{\beta}  |\tau^{[1]}_{\alpha} \rangle |\tau^{[2]}_{\alpha \beta} \rangle \ket{\omega^{[3,\cdots,N]}_\beta} \ .
\eeq
Proceeding iteratively as above for all the parties, the quantum state $\ket{\Psi}$ is finally expressed in terms of the Schmidt coefficients for different contiguous bipartitions as \footnote{This is indeed a matrix product state once $\ket{\tau^{[i]}_{\alpha \beta}}$ is expressed in terms of some local basis for each party $i = 1, \ldots, N$ \cite{guif1}.} 
\beq
\ket{\Psi}  = \sum_{\alpha, \beta , \ldots, \delta} \lambda^{[1]}_{\alpha} \lambda^{[2]}_{\beta}  \cdots  \lambda^{[N-1]}_{\delta} \ket{\tau^{[1]}_{\alpha}}\ket{\tau^{[2]}_{\alpha \beta}} \cdots \ket{\tau^{[N]}_{\delta}} \ .
\label{decomp}
\eeq

The above decomposition allows to obtain a useful expression for the fidelity $|\Lambda| = |\braket{\Phi}{\Psi}|$ between $\ket{\Psi}$ and some separable state  $\ket{\Phi} = \ket{\phi^{[1]}}  \ket{\phi^{[2]}} \cdots \ket{\phi^{[N]}}$ of the $N$ parties, in the limit $N \rightarrow \infty$ and for a translationally invariant state. To see this, let us previously define $D^{[i]}$ as the diagonal matrix of components  $D^{[i]}_{\alpha \beta} = \lambda^{[i]}_{\alpha} \delta_{\alpha \beta}$, and $M^{[i]}$ as the matrix of components $M^{[i]}_{\alpha \beta} = \braket{\phi^{[i]}}{\tau^{[i]}_{\alpha \beta}}$. In the limit $N \rightarrow \infty$ and for a system in one dimension invariant under translations divided in blocks (parties) of equal size $L$, we make three natural assumptions about the behavior of the physical system: $(i)$ it can be correctly described by a decomposition like the one in Eq.(\ref{decomp}), $(ii)$ its local description is site-independent, and $(iii)$ the maximization from Eq.(\ref{mma}) can be done with a state $\ket{\Phi}$ that is the tensor product of the same state $\ket{\phi}$ for all the parties. In this situation, we have that $\ket{\phi^{[i]}} = \ket{\phi}$, $\ket{\tau^{[i]}_{\alpha \beta}} = \ket{\tau_{\alpha \beta}}$, $M^{[i]} = M$ and $D^{[i]} = D$ for every party $i$. The fidelity $|\Lambda |$ is then given by
\beq
|\Lambda | = \lim_{N \rightarrow \infty}| d|^N  \ , 
\eeq
where $d$ is the eigenvalue of largest magnitude of the matrix $\sqrt{D} M \sqrt{D}$. Notice that the fidelity $|\Lambda |$ is zero for an infinite system unless it is a separable state of the blocks (see e.g. \cite{fid} and references therein). However, the global multipartite entanglement per block of size $L$ is finite and reads
\beq
\mathcal{E} = -\log{\left(|d_{{\rm max}}|^2\right)} \ ,
\eeq
where $|d_{{\rm max}}|$ is the maximum possible eigenvalue of matrix $\sqrt{D} M \sqrt{D}$ over all possible quantum states $\ket{\phi}$ such that $\braket{\phi}{\phi} = 1$ (which appear in the definition of matrix $M$). For a given state $\ket{\phi}$, the maximum eigenvalue of matrix $\sqrt{D} M \sqrt{D}$ can be obtained by solving a different maximization problem, namely, 
\beq
|d|^2 = \left(\argmax_{\vec{r}} \left|\vec{r}^{\ \dagger} \ \sqrt{D} M \sqrt{D} \ \vec{r}\  \right| \right)^2\ ,
\label{maxi}
\eeq
where $\vec{r}^{\ \dagger} \vec{r} = 1$. Therefore, to find $|d_{{\rm max}}|$ we need to solve two different maximization problems: one over the quantum states $\ket{\phi}$ and another over the vectors $\vec{r}$.

\begin{figure}[h]
\includegraphics[width=0.48\textwidth]{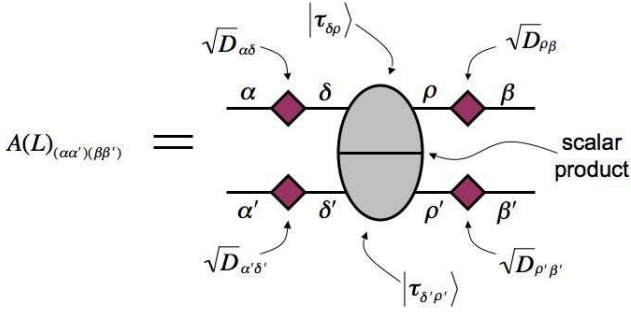}
\caption{(color online) Diagrammatic representation of the components $A(L)_{(\alpha \alpha'),(\beta \beta')}$ of matrix $A(L)$ for a block of size $L$. In the diagram, the violet diamonds correspond to the components $\sqrt{D}_{\alpha \beta}$ of matrix $\sqrt{D}$ and the half-ellipses correspond to quantum states $\ket{\tau_{\alpha \beta}}$. The scalar product between two quantum states is represented by two half-ellipses together, one of them representing the {\it bra} $\bra{ \ }$ and the other the {\it ket} $\ket{ \ }$. The different emergent lines correspond to the Greek indices $\alpha,\beta...$, where common legs between objects correspond to shared summed indices, and free legs correspond to free indices.}
\label{fig2}
\end{figure}

Our interest is now focused on solving this double maximization problem. In order to achieve this, we fix vector $\vec{r}$ and perform the maximization over the quantum state $\ket{\phi}$ only. As a result, we obtain that the optimal state $\ket{\phi_{\rm max}}$ is given by $\ket{\phi_{\rm max}} = \ket{\psi(\vec{r})}/\sqrt{\braket{\psi(\vec{r})}{\psi(\vec{r})}}$, with $\ket{\psi(\vec{r})} \equiv \sum_{\alpha, \beta} r^*_{\alpha} \sqrt{\lambda_{\alpha}}r_{\beta} \sqrt{\lambda_{\beta}}\ket{\tau_{\alpha \beta}}$. We then have that
\beqa
|d_{{\rm max}}|^2 &=& \argmax_{\vec{r}} \left|\braket{\psi(\vec{r})}{\psi(\vec{r})}\right| \nonumber \\
&=& \argmax_{\vec{r}} \left|(\vec{r} \otimes \vec{r}^{\ *})^{\dagger} A(L)(\vec{r} \otimes \vec{r}^{\ *})\right| \ .
\label{maxom}
\eeqa
In the above expression we have introduced matrix $A(L)$ for a block of size $L$. The components $A(L)_{(\alpha \alpha'),(\beta \beta')}$ of this matrix (where $(\alpha \alpha')$ is understood as a single index and similarly for $(\beta \beta')$) are computed as shown in the diagram of Fig.(\ref{fig2}). 

\begin{figure}[h]
\includegraphics[width=0.45\textwidth]{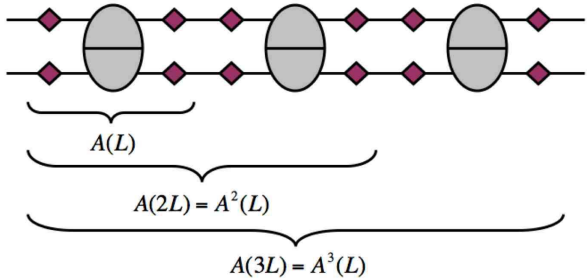}
\caption{(color online) The repeated multiplication of matrix $A(L)$ for a block of size $L$ produces the same matrix but for blocks of larger size, i.e. $A(kL) = A^k(L)$, for $k = 2,3...$.}
\label{fig3}
\end{figure}

Before moving to the maximization of Eq.(\ref{maxom}), we wish to explain certain spectral properties of matrix $A(L)$ that turn out to be of relevance for our purposes. First of all, notice that $A(kL) = A^k(L)$ for $k = 2,3...$, see Fig.(\ref{fig3}). The reason for this is that a block of size $2L$ can be understood by grouping together two contiguous blocks of size $L$, which at the level of the matrix involves the multiplication of $A(L)$ by itself. Moreover, we have the chain of equalities $1 = \braket{\Psi}{\Psi} = \lim_{k \rightarrow \infty} {\rm tr} \left(A(L)^k\right) =\lim_{k \rightarrow \infty} \left(\nu_1(A(L))\right)^k$, where $\nu_1(A(L))$ is the eigenvalue of largest magnitude of $A(L)$. These relations necessarily imply that $\nu_1(A(L)) = 1$ and $|\nu_i(A(L))| < 1$ for the rest of the non-zero eigenvalues $\nu_i(A(L))$, $i = 2,3...$ of $A(L)$. In fact, there is only one eigenvalue of magnitude $1$ in the case of systems away from criticality. This is so since the correlation length $\xi$ of a one-dimensional quantum system in state $\ket{\Psi}$ is given by $(\xi/\epsilon) = -1/\log{|\nu_2(A(\epsilon))|}$, where $\epsilon$ is a cut-off parameter at short distances that coincides with the lattice spacing in the case of systems defined on a lattice (see e.g. \cite{scholl} for a derivation of this property).

As a consequence of the above facts, a key property of matrix $A(L)$ can be derived: the spectral decomposition of $A(L)$ reads 
\beq
A(L) = \sum_{i}\left(\nu_i(A(\epsilon))\right)^{(L/\epsilon)} \vec{a}(i) \vec{a}(i)^{\dagger} \ ,
\label{spectral}
\eeq
where $\vec{a}(i)$ is the eigenvector corresponding to the $i$th eigenvalue of largest magnitude. From this equation it is possible to see that, if $L \gg \xi$, then all the eigenvalues for $i > 1$ are exponentially suppressed as $exp(-(L/\xi_i))$ with $(\xi_i/\epsilon) = \log{|\nu_i(A(\epsilon))|}$, and therefore $A(L) \sim \vec{a}(1) \vec{a}(1)^{\dagger}$ since only the eigenvalue of largest magnitude $\nu_1(A(\epsilon)) = 1$ does not vanish. Under these circumstances, the maximization from Eq.(\ref{maxom}) transforms into
\beq
|d_{{\rm max}}|^2 =\left(\argmax_{\vec{r}} \left|(\vec{r} \otimes \vec{r}^{\ *})^{\dagger}\vec{a}(1)\right|\right)^2 \ ,
\label{maxum}
\eeq
which is the maximization of a scalar product between two vectors. 

In order to find $|d_{{\rm max}}|$ in Eq.(\ref{maxum}), we make use of some properties of the decomposition  in Eq.(\ref{decomp}) of state $\ket{\Psi}$. In particular, we use the fact that the corresponding eigenvector $\vec{a}(1)$ has components $a(1)_{\alpha \alpha'} = \lambda_{\alpha} \delta_{\alpha \alpha'}$, which is a consequence of the orthonormalization of the different left and right Schmidt vectors in Eq.(\ref{decomp}) \footnote{In fact, Eq.(\ref{decomp}) is a matrix product state in canonical form \cite{OV}, from which the property immediately follows.}. The optimization from Eq.(\ref{maxum}) then gives  $|d_{{\rm max}}|^2 = \lambda_1^2$, where $\lambda_1$ is the largest Schmidt coefficient in the decomposition of $\ket{\Psi}$ given in Eq.(\ref{decomp}). The global geometric entanglement per block of size $L$ is then given by
\beq
\mathcal{E} = -2 \log{(\lambda_1)} \ ,
\label{ee}
\eeq
which holds away from criticality for $L \gg \xi$. 

At this point the global entanglement $\mathcal{E}$ only depends on the largest Schmidt coefficient $\lambda_1$. However, it is possible to obtain a more convenient expression for $\mathcal{E}$ in terms of the correlation length $\xi$. This can be achieved by considering the reduced density matrix $\rho$ of one of the blocks, given by 
\beq
\rho = \sum_{\alpha,\beta} \left(\lambda_{\alpha} \lambda_{\beta} \right)^2 \ket{\tau_{\alpha \beta}} \bra{\tau_{\alpha \beta}} \ .
\label{ro}
\eeq
(see e.g. \cite{OV} for details on this derivation). Remarkably, the following inequality holds: 
\beq 
(\nu_1(\rho))^{1/4} \ge \lambda_1 \ge (\nu_2(\rho) + \nu_{n}(\rho))^{1/4} \ ,
\label{desi}
\eeq
where $\nu_i(\rho)$ refers to the $i$th largest eigenvalue of the reduced density matrix $\rho$, and $\nu_{n}(\rho)$ refers to the smallest one \footnote{Here, $(\nu_1(\rho))^{1/4} \ge \lambda_1$ holds since $(\lambda_{\alpha} \lambda_{\beta})^2$ in Eq.(\ref{ro}) are the probabilities of the statistical mixture of states $\ket{\tau_{\alpha \beta}}$ in $\rho$, and using Theorem $10$ in \cite{NV} the result follows. 
The second part of Eq.(\ref{desi}) can be proven by using Weyl's inequalities \cite{bathia}: we obtain that $\lambda_1^4 = \nu_1(\lambda_1^4 \ket{\tau_{11}} \bra{\tau_{11}}) = \nu_1(\rho - \sigma) \ge \nu_{n}(\rho) + \nu_1(-\sigma)$, where $\sigma = \rho - \lambda_1^4 \ket{\tau_{11}} \bra{\tau_{11}}$ where we used a suitable Weyl inequality. Then, $\nu_1(-\sigma) = \nu_1(\lambda_1^4 \ket{\tau_{11}} \bra{\tau_{11}} - \rho) \ge \nu_{n - 1}(-\rho) = \nu_2(\rho)$, using again a Weyl inequality. From here, we get the desired relation.}. Given the fast decay of the eigenvalues of $\rho$ away from criticality (see e.g. the second reference in \cite{fin1} for a discussion about this property), we can safely assume that $\nu_{n}(\rho)/\nu_{2}(\rho) \sim 0$. Also, following the results from Sec.IV in \cite{ent3}, it is not difficult to see that $-\log{(\nu_2(\rho))} \sim -\log{(\nu_1(\rho))} \sim (c/6) \log{(\xi/\epsilon)}$ if the system is close enough to a quantum critical point with central charge $c$ (so that $\xi$ is large), which implies our claim in Eq.(\ref{equ2}) in the introduction. 

{\it Upper bound for critical systems.-}
At a quantum critical point, the correlation length $\xi$ of the system diverges, so that several of the eigenvalues of matrix $A(L)$ in Eq.(\ref{spectral}) are expected to be of magnitude one and not necessarily real. This, in turn, makes the maximization of Eq.(\ref{maxom}) quite difficult. However, it is still possible to derive a general upper bound on the scaling of $\mathcal{E}$ with the size $L$ of the blocks.

This upper bound is derived as follows: for a finite system of $M$ parties, the squared fidelity $|\Lambda|^2$ between a mixed state $\rho_M$ for the $M$ parties and a separable product state $\ket{\Phi} = \ket{\phi}^{\otimes N}$ is given by $|\Lambda|^2 =  \bra{\Phi} \rho_N\ket{\Phi}$. We can assume that the reduced density matrix $\rho_M$ describes the degrees of freedom of $M$ contiguous blocks in a pure quantum state of $N > M$ parties, which is described by the decomposition in Eq.(\ref{decomp}). The density matrix $\rho_M$ is then obtained by tracing out the degrees of freedom of the remaining $N-M$ sites in Eq.(\ref{decomp}). If the whole system is translationally invariant, and for infinite $N$, it is possible to define a site-independent vector $\vec{b}$ of components $b_{(\alpha \alpha')} = \sqrt{\lambda_{\alpha}} M_{\alpha \beta}\sqrt{\lambda_{\alpha'}} M_{\alpha' \beta}^* \lambda_{\beta}^2$ (where matrix $M$ is defined as in the off-critical case). The squared fidelity $|\Lambda|^2$ is then expressed in terms of vector $\vec{b}$ as $|\Lambda|^2 = \vec{b}^{\ \dagger} \left( A(L) \right)^{M-2} \vec{b}$, where matrix $A(L)$ is defined as previously. Next, we make use of the normalized vector $\vec{a}(1)$ of components $\vec{a}(1)_{(\alpha \alpha')} = \lambda_{\alpha} \delta_{\alpha \alpha'}$. Introducing a resolution of the identity operator before and after each one of the matrices $A(L)$ in the expression for $|\Lambda|^2$, and in such a way that the resolution includes the projector $\vec{a}(1) \vec{a}(1)^{\dagger}$, we obtain $|\Lambda|^2 = \left(\vec{b}^{\ \dagger} \vec{a}(1) \right)^{M} + \Upsilon$, where $\Upsilon$ is some positive constant. It is possible to see that this expression is indeed equivalent to
\beq
|\Lambda|^2 = \left(\bra{\phi} \rho \ket{\phi} \right)^M + \Upsilon \ ,
\label{deki}
\eeq
where $\rho$ is the reduced density matrix of a block of size $L$. From the above relation, we derive after some manipulation the inequality $|\Lambda_{\rm max}|^2 > (\nu_1(\rho))^M$ for the maximum overlap $|\Lambda_{\rm max}|^2$, where $\nu_1(\rho)$ is the largest eigenvalue of $\rho$. The global geometric entanglement $\mathcal{E}$ per block  is then seen to be bounded in the limit $M \rightarrow \infty$ as
\beq
\mathcal{E} < -\log{(\nu_1(\rho))} = E_1(\rho) \ , 
\label{upbound}
\eeq
where $E_1(\rho)$ is the single-copy entanglement between a block of size $L$ and the rest of the system. Now we make use of the existing results on the behaviour of $E_1(\rho)$ at the critical point of a quantum phase transition from \cite{single}. In particular, we use the property that, at criticality, the single-copy entanglement $E_1(\rho)$ scales with the size $L$ as  $E_1(\rho) = (c/6) \log{(L/\epsilon)}$ for large $L$, where $c$ is the central charge of the underlying conformal field theory in $(1+1)$ dimensions. By combining this result with Eq.(\ref{upbound}), the expression from Eq.(\ref{equ3}) in the introduction  follows, providing a universal upper bound on $\mathcal{E}$ for any critical quantum system in one spatial dimension in terms of a logarithmic function of the size $L$ of the blocks. 

{\it Conclusions.-} Here we have established first analytical derivations of the global geometric entanglement per block of size $L$ for quantum systems in one spatial dimension and invariant under translations. We have proven that one-dimensional quantum systems tend to be globally separable along RG flows by following a universal scaling law in the correlation length $\xi$ of the system. Furthermore, an upper bound on the critical scaling of the global geometric entanglement has been given in terms of the logarithm of the size $L$. Our results are analytical, universal, and are valid for all one-dimensional quantum systems close to and at criticality. 

{\it Acknowledgements.-} We acknowledge discussions with L. Tagliacozzo and G. Vidal. 

{}
\end{document}